\documentclass[nofootinbib, amsmath,amssymb, aps,twocolumn]{revtex4}

\usepackage{hyperref}

\usepackage{graphicx}% Include figure files
\usepackage{dcolumn}% Align table columns on decimal point
\usepackage{bm}% bold math

\usepackage{booktabs}
\usepackage{topcapt}

\usepackage{xcolor}

\usepackage{slashed} 

 \usepackage{axodraw2}
 
 \usepackage{ulem}
 
\begin{document}

\title{Deconfinement, Center Symmetry and the Ghost Propagator in Landau Gauge Pure SU(3) Yang-Mills Theory}% Force line breaks with \\

\author{V\'{\i}tor Paiva}
\email{vpaiva462@gmail.com}
\author{Paulo J. Silva}
\email{psilva@uc.pt}
\author{Orlando Oliveira}
\email{orlando@uc.pt}
\affiliation{CFisUC, Departament of Physics, University of Coimbra, 3004-516 Coimbra, Portugal}

\date{\today}% It is always \today, today,
             %  but any date may be explicitly specified

\begin{abstract}
The temperature dependence of the Landau gauge ghost propagator is investigated in pure SU(3) Yang-Mills theory with lattice QCD simulations.
Its behavior around the confined-deconfined phase transition temperature, $T_c \sim 270$ MeV, is investigated. 
The simulations show that in the deconfined phase, the ghost propagator is  enhanced for small momenta, $\lesssim 1$ GeV.
Furthermore, the analysis of the spontaneous breaking of center symmetry on the ghost propagator is studied. 
Similarly as observed for the gluon propagator, the simulations result in a decoupling of the sectors where the phase of the Polyakov loop is either
0 or $\pm 2\pi/3$ sectors, with the latter remaining indistinguishable. 
The results point to the possible use of the ghost propagator as an "order parameter" for the confined-deconfined phase transition.
\end{abstract}

%\keywords{Suggested keywords}%Use showkeys class option if keyword
                              %display desired
\maketitle

\tableofcontents

%====================================================================
\section{Introduction and Motivation}

The dynamics of quarks and gluons feels the  effects of the surrounding hadronic environment that
translates into a rich QCD phase diagram, see e.g. \cite{Fukushima:2010bq,Andersen:2014xxa,Odyniec:2022ztn} 
and citations therein. 
The QCD phase diagram is accessed through the Green functions and, therefore, the fundamental
propagators and vertices are not blind to temperature $T$, density $\rho$, external magnetic fields or any other field
felt by the fundamental QCD quanta. 
For example, at zero hadronic density, from the quark propagator one can compute the running quark mass that
shows different patterns at low and high temperatures. At zero and low temperatures the running quark mass is momentum
independent, illustrating  the chiral symmetry breaking mechanism \cite{Oliveira:2018lln,Virgili:2022wfx}.
On the other hand, for sufficiently high temperatures, (quenched) lattice results for the quark propagator suggest that chiral symmetry is restored
and that quarks behave essentially as free particles \cite{Oliveira:2019erx}. These results suggest that as 
the hadronic temperature increases, quarks go from a confined phase, where they appear as constituents of hadrons, to a deconfined phase
with the formation of a strongly  interacting  plasma of quarks and gluons. Moreover, the studies of the finite temperature gluon propagator also show an equivalent transition for the same range of temperatures --- see, for example, \cite{Silva:2013maa}. This change implies a modification of the analytical properties of the propagator that has been foreseen
by several studies.

For full QCD, the temperature where the transition between phases occurs has been estimated to be $T_c \sim 150$ MeV \cite{Borsanyi:2010bp,Bazavov:2011nk}. 
 The investigation of full QCD is demanding
 and, alternatively, the study of pure Yang-Mills theory, which is less complex due to the dismissal of quark contributions to the dynamics, can help understand the behavior of hadronic matter at finite temperature. The main differences with
respect to full QCD are the nature of the order of the phase transition and that gluons become deconfined at higher temperatures.
For pure SU(3) Yang-Mills the confined-deconfined transition is first order and the critical temperature is
 $T_c \sim 270$ MeV.
 
For pure SU(3) Yang-Mills theory, the transition from low temperatures to higher temperatures can be observed by looking at the
gluon propagator, see e.g. \cite{Dumitru:2012fw,Maas:2011ez,Aouane:2011fv,Bornyakov:2011jm,Maas:2011se,Aouane:2012bk,Silva:2013maa,Falcao:2020vyr,Siringo:2021fxo,vanEgmond:2022nuo}
and references therein. At zero temperature the gluon propagator in the minimal Landau gauge
is finite for all range of momenta. The theory generates dynamically a mass scale, and the propagator exhibits positivity violation that 
can be viewed as an indication of gluon confinement \cite{Cornwall:2013zra,Li:2019hyv,Oliveira:2022rwu}.
At finite temperature the gluon propagator has an electric and a magnetic form factor, and it is common to associate a mass scale with each
component. For temperatures above $T_c$, these electric and magnetic masses scale differently with temperature \cite{Kapusta:2006pm}.
From the point of view of the confined-to-deconfined phase transition, it was observed that the electric form factor can be used to
identify the transition and its nature \cite{Maas:2011ez,Silva:2013maa,Silva:2016onh}. Moreover, for sufficiently high temperatures the
gluons behave as quasi-particles with a finite mass that grows with $T$. The behaviour of the ghost propagator below and above $T_c$ has been
touched in \cite{Aouane:2012bk}, with the lattice QCD results suggesting an enhancement of the IR propagator that grows with $T$, 
but a systematic study and specially the role of the center symmetry is still missing in the literature. The current work
aims to fill this gap and investigate these issues close to the critical temperature associated with the confined-to-deconfined transition.

In QCD, like in any gauge theory, the set of gauge related configurations, i.e. the configurations on a gauge orbit, are equivalent and express
the same physical content. Choosing a gauge means identifying a gauge configuration on each gauge orbit. This represents a problem not yet
solved in the quantization of Yang-Mills theories. Despite the gauge fixing procedure being fundamental to handle gauge fields, given the 
complexity involved, in this paper we will not discuss  the issues related to choice of a gauge. 
We just recall the reader that how gauge fixing is implemented can impact on the IR propagators
\cite{Cucchieri:1997dx,Silva:2007tt,Silva:2010vx,Silva:2004bv,Sternbeck:2012mf}.

For pure Yang-Mills SU(N) theory the order parameter for the confined-to-deconfined transition is the Polyakov loop that in the continuum formulation
and in Euclidean space-time is defined as
\begin{equation}
L(\vec{x}) = \frac{1}{N} \mbox{Tr} \left\{ \mathcal{P} \, \exp \left[  i \, g \, \int^{1/T}_0 dx_4 \, A_4( \vec{x}, x_4)  \right] \right\} \ ,
\end{equation}
where $\mathcal{P}$ means path ordering and $T$ is the temperature. The space average of the Polyakov loop
\begin{equation}
   L = \langle L(\vec{x}) \rangle_{\vec{x}} \, \propto \, e^{- F_q / T}
\end{equation}
measures the free energy of a static quark $F_q$, see e.g. \cite{Kapusta:2006pm}.
In the confined phase where $T < T_c$ the quark free energy is infinite and it follows that the renormalized Polyakov loop is $L = 0$.
For temperatures above $T_c$ the renormalized space averaged Polyakov loop is one \cite{Lo:2013etb}; it follows that $F_q = 0$  , meaning that quarks behave essentially as free particles. The Polyakov loop is associated with the gluon content of the theory.
On a finite hypercubic $N^3_x \times N_t$ lattice the bare Polyakov loop is given by
\begin{equation}
L(\vec{x}) = \prod\limits_{x_4 = 0}^{N_t} ~ \mathcal{U}_4 (\vec{x}, x_4)
\end{equation}
where $\mathcal{U}_4$ is the time-oriented link.

For pure Yang-Mills theory the generating functional and, therefore, the Green functions
have an additional symmetry associated with an invariance under global gauge transformations of the center group.
For $N = 3$ the center group is
\begin{equation}
Z_3 = \left\{ \, 1, \, e^{i \, 2 \, \pi / 3}, \, e^{- \, i \, 2 \, \pi / 3} \, \right\}
\end{equation}
and divide the group SU(3) into equivalent classes.
A $Z_3$ global transformation leaves  the action unchanged, but the Polyakov loop $L$ acquires an extra phase. Note that this invariance applies  for both the continuum and 
lattice formulations.  The inclusion of fermions breaks the center symmetry
and can change the nature of the phase transition, see e.g. \cite{Kaiser:2022vjg} and references therein.

Our lattice study for the pure gauge SU(3) Yang-Mills theory uses the Wilson action. To explore the center symmetry and, in particular,
the breaking of the center symmetry for $T > T_c$, one considers the transformation where the links on a given hyperplane chosen
at $x_4 = const$ are multiplied by $z \in Z_3$.  Under this transformation, the Polyakov loop $L(\vec{x})$ becomes $z \, L(\vec{x})$.
In the confined phase, $L = 0$ and the space averaged Polyakov loop is invariant under center transformations. 
However, for temperatures above $T_c$, where $L \ne 0$, the Polyakov loop phase is changed by the center transformation and, in this way,
one can go through the various equivalent classes and compute, for each class of gauge configurations, the propagator. In this way one can 
investigate how the breaking of the center symmetry impacts the fundamental fields.

For the minimal Landau gauge and the SU(3) group,
the gluon propagator was studied in \cite{Silva:2016onh} below and above the critical temperature. 
The authors showed that for temperatures below $T_c$ the gluon propagator is the same, within the statistical precision achieved in the
simulation, for  all the equivalent classes. For $T > T_c$ it was found that the gluon propagator differs, mainly at low momenta, between
the class of configurations associated with a vanishing phase of the Polyakov loop, let us call it the zero  sector, and the other
two sectors where the phase of $L$ is $\pm 2 \pi /3$. Then, for temperatures above the critical temperature the center symmetry is
spontaneously broken. Moreover, this difference can be used as an order parameter to identify the transition to the deconfined phase.

Herein, we pursue and extend the work of \cite{Silva:2016onh} to the ghost sector. The details of the sampling, gauge fixing, of the mapping
and identification of each of the equivalent classes of gauge configurations can be read in the above cited paper. The computation of ghost
sector requires the inversion of a large sparse matrix and, therefore, it is computationally demanding. For this reason, only a subset of the 
gauge configurations, for temperatures above and below $T_c$, considered in \cite{Silva:2016onh} will be investigated for each of the
equivalent classes. As described below we observe that  above $T_c$, the ghost propagator is enhanced in the infrared region and,
as for the gluon propagator, the ghost propagator in the zero  sector differs from the ghost propagator in the two equivalent classes where $L$
acquires a non-vanishing phase. Indeed, the low momentum ghost propagator in the zero  sector is enhanced when compared  to
the other equivalent classes. Recall that for the gluon propagator, the transverse (electric) form factor is enhanced in the zero 
sector, relative to the other equivalent classes, while the longitudinal (magnetic) form factor associated with the zero  sector is suppressed 
relative to the remaining  sectors, see \cite{Silva:2016onh}. The observed correlation between the ghost and the gluon electric and magnetic form
factors remains to be understood theoretically.  As for the gluon propagator,
we observe that  the differences seen in the ghost propagator computed for the different $Z_3$ sectors can be used to identify the 
confined-to-deconfined transition and, therefore, the spontaneous symmetry breaking of the center symmetry.
Preliminary results can be found in \cite{Oliveira:2022jir}.

The current work is organised as follows. In  Sec. \ref{Sec:Setupt}, the lattice setup is outlined, focusing on the procedures set for the statistical 
analysis of the Monte Carlo simulations, the elimination of lattice artefacts and data renormalization. In  Sec. \ref{sec:tempghost}, the effects of temperature on 
the ghost propagator are discussed. In Sec. \ref{Sec:Ghost2}, the effects of the spontaneous center symmetry breaking on the ghost propagator are analyzed. 
Finally, in Sec. \ref{Sec:fim}, a summary of the main results and a direction toward future studies are presented.

%====================================================================
%====================================================================
\section{Lattice Setup and the Landau Gauge Ghost Propagator \label{Sec:Setupt}}

The QCD simulations reported here use  lattices $N^3_s\times N_t$, with $N_t \ll N_s$. The temperature is defined as $T = 1/N_t \, a$, where $a$ is the lattice spacing. We use the Wilson gauge action to perform the importance sampling.  The ghost propagator in momentum space
is given by
\begin{equation}
   G^{ab} (p) = - \, \delta^{ab} \, G(p^2) =  - \, \delta^{ab} \, \frac{d_g(p^2)}{p^2} \ ,
\end{equation}
where the latin letters refer to color indices, $G(p^2)$ is the ghost propagator and $d_g(p^2)$ is the ghost dressing function.
On the lattice, the inverse of the ghost propagator is given by the second variation of the gauge fixing functional that results on
a large sparse matrix. This matrix has zero modes and for the computation of ghost propagator, its inversion is performed using the conjugate
gradient method following \cite{Suman:1995zg}. All computer simulations have been done with the help of Chroma \cite{chroma} and PFFT \cite{pfft} libraries.

\begin{table}[t]
\centering
\begin{tabular}{c@{\hspace{0.5cm}}c@{\hspace{0.5cm}}lr@{\hspace{0.5cm}}l@{\hspace{0.5cm}}l}
\hline
$T$     &    $\beta$ & $N_s$ &  $N_t$ &  $a$  & $1/a$ \\
(MeV)  &                &             &             &  (fm)  &  (GeV) \\
\hline
121 &    6.0000 & 64    &     16 &     0.1016 &     1.943 \\
%162 &    6.0000 & 64    &     12 &     0.1016 &     1.943 \\
194 &    6.0000 & 64    &     10 &     0.1016 &     1.943 \\
243 &    6.0000 & 64    &     8 &     0.1016 &     1.943 \\
260 &    6.0347 & 68    &     8 &     0.09502 &     2.0767 \\
265 &    5.8876 & 52    &     6 &     0.1243 &     1.5881 \\
275 &    6.0684 & 72    &     8 &     0.08974 &     2.1989 \\
%285 &    5.9266 & 56    &     6 &     0.1154 &     1.7103 \\
%290 &    6.1009 & 76    &     8 &     0.08502 &     2.3211 \\
%305 &    5.9640 & 60    &     6 &     0.1077  &      1.8324 \\
%305 &    6.1326 & 80    &     8 &     0.08077 &     2.4432 \\
324 &    6.0000 & 64    &     6 &     0.1016  &      1.943 \\
366 &    6.0684 & 72    &     6 &     0.08974 &     2.1989 \\
%397 &    5.8876 & 52    &     4 &     0.1243 &     1.5881 \\
%428 &    5.9266 & 56    &     4 &     0.1154 &     1.7103 \\
%458 &    5.9640 & 60    &     4 &     0.1077  &      1.8324 \\
486 &    6.0000 & 64    &     4 &     0.1016  &      1.943 \\
\hline
\end{tabular}
\caption{Lattice setup. Each $T$ uses 100 gauge configurations and, for each inversion to compute the ghost propagator, 
two point sources were averaged. See main text for details.}
\label{tab-1}
\end{table}

The lattice setup used in the investigation of the ghost sector reported in Section \ref{sec:tempghost} is summarized in Tab. \ref{tab-1}. For each temperature, 100 gauge configurations
rotated to the Landau gauge were considered and, in the computation of the ghost propagator we used two point sources
to compute point-to-all propagators. As point sources we took the lattice origin $(0, \, 0, \, 0, \, 0)$ and the lattice spatial mid-point 
$(N_s/2, \, N_s/2, \, N_s/2, \, 0)$. The corresponding momentum propagators for the two point sources were averaged, and the average used in the
statistical analysis of the Monte Carlo simulation. As in previous works, herein only the results coming from the first Matsubara frequency are reported.

The formulation of the gauge theory on a finite hypercubic lattice breaks rotational symmetry and introduces lattice artefacts. In order to 
minimise the lattice artefacts, the large momenta, defined as the physical momenta above 1 GeV, were subject of a cylindrical cut 
\cite{Leinweber:1998uu}. The momenta cut chooses only the momenta whose distance, $d$, from the lattice diagonal, is such that 
$d\, a < 4 \,(2\pi/N_s)$, i.e. it considers only momenta less than four spatial units away from the lattice's diagonal, ($p$, $p$, $p$, 0).

In order to be able to compare the results of simulations performed with different lattice spacings, the ghost data has to be renormalized. 
Herein, the renormalization was performed at $\mu=4$ GeV, demanding that $G(\mu^2)=1/\mu^2$. 
In the computation of the renormalization constants $Z$, for each simulation the lattice data was fitted to the functional form
\begin{equation}
    \label{eqn-1}
    G_{fit}(p^2)=\frac{b \, + \, c \, p^2}{p^4 \, + \, d \, p^2 \, + \, e} \ ,
\end{equation}
 where $b$, $c$, $d$ and $e$ are adjustable parameters, and then
 \begin{equation}
    Z = \frac{1}{\mu^2 ~ G_{fit}(\mu^2)} \ .
\end{equation}    
The fits were performed for momenta in the range $[ 3 \, , \, 5]$ GeV and the associated $\chi^2/d.o.f. \sim 1$.

%====================================================================
%====================================================================
\section{Ghost Propagator --- Temperature effects}
\label{sec:tempghost}

In this section we report on the dependence of the Landau gauge ghost propagator with temperature. The configurations used are
such that the phase of the Polyakov loop vanishes. As described previously, this can be a\-chie\-ved exploring the center symmetry transformations.
Moreover, the lattice data is described as a function of the improved lattice momentum
\begin{eqnarray}
q_\mu & = & \frac{2}{a} \, \sin \left( \frac{\pi}{N_\mu} \, n_\mu\right) \ , \nonumber \\
 \qquad n_\mu & = & - \frac{N_\mu}{2}, \, \dots, \, -1, \, 0, \, 1, \, \dots , \, \frac{N_\mu}{2}-1 \ ,
\end{eqnarray}
where $N_\mu$ is the number of lattice points in direction $\mu$, $a$ is the lattice spacing, and not as a function of the lattice naive momentum
\begin{equation}
p_\mu  =  \frac{2 \, \pi}{a \, N_\mu} \, n_\mu \ .
\end{equation}
The motivation for using $q_\mu$ instead of $p_\mu$ comes from lattice perturbation theory. Recall that for bosonic fields the propagator is a 
function of $q^2$ and not of $p^2$, see e.g. \cite{Montvay:1994cy}, and also because is was observed that using $q$ instead of $p$ reduce
the lattice artefacts, see \cite{Leinweber:1998uu,Becirevic:1999uc,Silva:2004bv,deSoto:2007ht,Vujinovic:2018nqc,Catumba:2021hcx}.

\begin{figure*}[t] %  figure placement: here, top, bottom, or page
   \centering
   \includegraphics[width=3.47in]{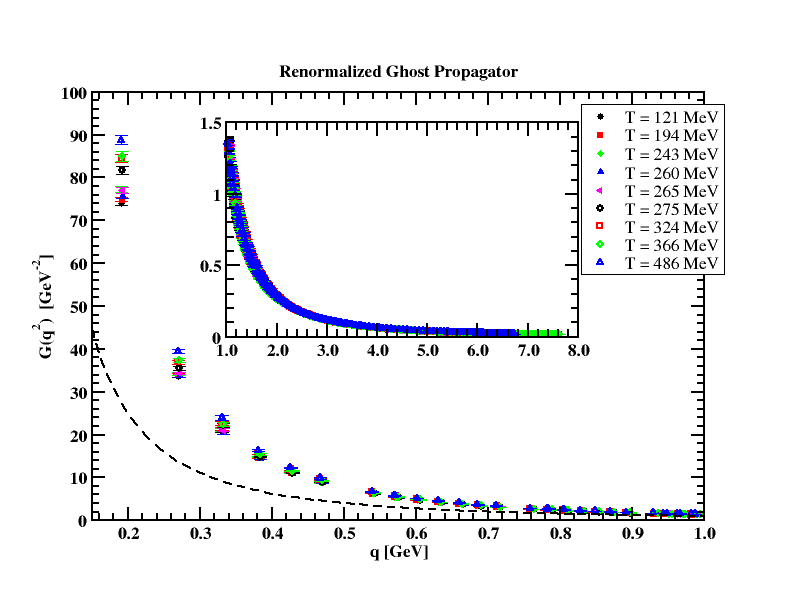} ~
   \includegraphics[width=3.47in]{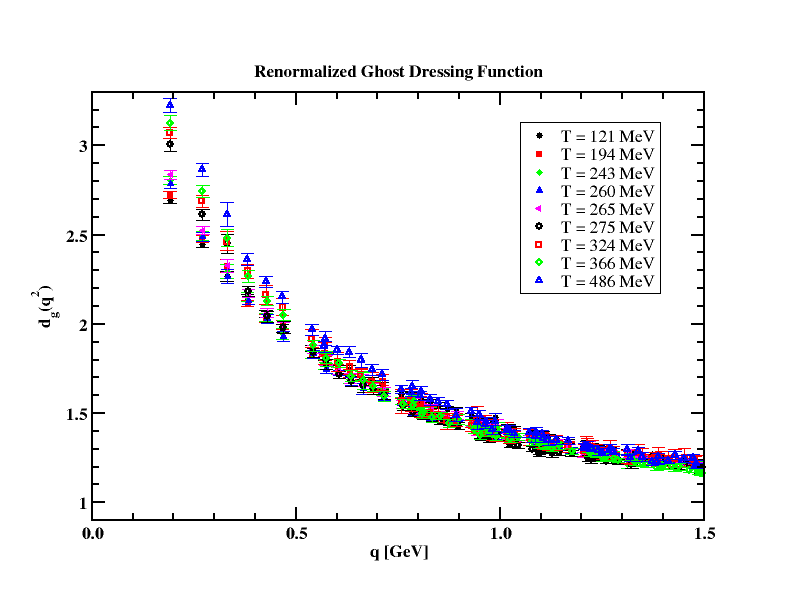}   
   \caption{Landau gauge ghost propagator as a function of the improved momenta $q$ for various temperatures and for the first Matsubara frequency. 
                 The left plot shows the propagator for various $T$, with the dashed line representing the tree level perturbative ghost propagator, 
                 while the right plot reports the ghost dressing function.}
   \label{fig:ghostT}
\end{figure*}

The Landau gauge ghost propagator and dressing function for various $T$ can be seen in Fig. \ref{fig:ghostT}.
As a first comment to the results we notice that the qualitative behavior of the ghost propagator at finite $T$ 
follows the momentum dependence observed at $T = 0$, see e.g. 
\cite{Cucchieri:2007md,Cucchieri:2007zm,Bogolubsky:2009dc,Ilgenfritz:2010gu,Duarte:2016iko} and references therein.
In the infrared region, $G(p^2)$ diverges with a power greater than the prediction of perturbation theory as can be observed in
the right hand plot of Fig. \ref{fig:ghostT}. For zero temperature there are several studies that support a pole at zero momentum for the lattice
ghost propagator \cite{Falcao:2020vyr,Dudal:2019gvn,Boito:2022rad}, and the data reported in Fig. \ref{fig:ghostT} suggests that
the same should occur for finite $T$. Furthermore, by fitting the lattice Landau ghost propagator at the highest available momenta, in all
cases, the lattice data is well reproduced by the tree level perturbation result for $p \gtrsim 5$ GeV. Indeed, for the fits of the lattice data 
to
\begin{equation}
   G(p^2) = \frac{Z}{p^2}
\end{equation}
it follows that $\chi^2/d.o.f. \sim 1$ and the corresponding residua are $Z \sim 1$. Certainly, the inclusion of the leading log will improve
the range of momenta where the matching between lattice simulations and perturbation theory occurs, see e.g. \cite{Duarte:2016ieu}.
We take this result as an indication that, in the ultraviolet region, the lattice data reproduces the prediction of perturbation theory.

\begin{figure}[t] %  figure placement: here, top, bottom, or page
   \centering
   \includegraphics[width=3.4in]{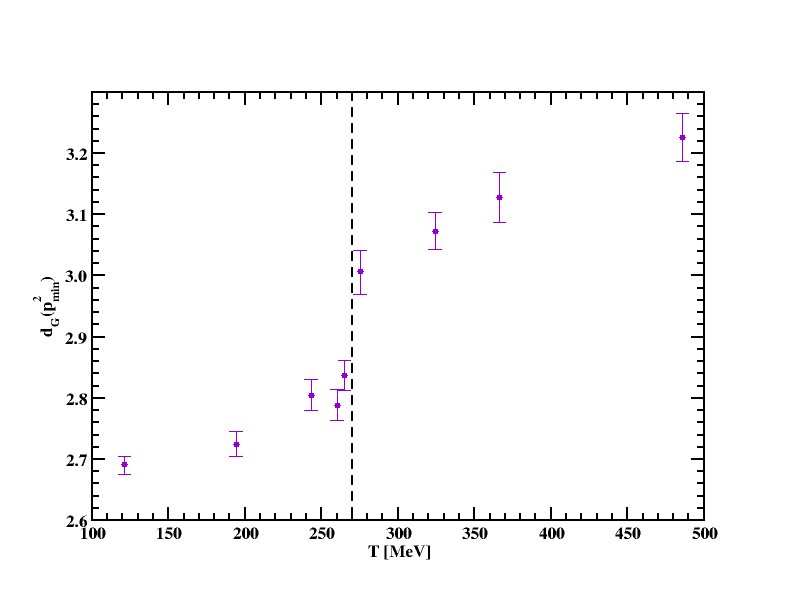} 
   \caption{Landau gauge ghost dressing function at the lowest available momenta $q \sim 0.19$ GeV as a function of $T$.}
   \label{fig:2}
\end{figure}

The second comment concerns the temperature dependence of $G(p^2)$ or, equivalently, of $d_G(p^2)$. As can be observed in Fig. \ref{fig:ghostT}
the ghost propagator at low momenta increases when the temperature is increased. The enhancement of the infrared $G(p^2)$ is stronger when 
the confinement-to-deconfinement temperature is crossed that for pure Yang-Mills takes the value $T_c  \sim 270$ MeV, see 
\cite{Iwasaki:1992ik,Boyd:1996bx,Lucini:2001ej,Silva:2016onh}
and references therein. Note that in  Fig. \ref{fig:ghostT} the results of the simulations for $T > T_c$ are represented by open symbols.
To illustrate the stronger enhancement that occurs around $T \sim T_c$, in Fig. \ref{fig:2} we show the ghost dressing function for the
smallest momenta available in each simulation, $q \sim 0.19$ GeV. The curve mimics the behavior observed for the temperature
dependence of the longitudinal gluon propagator \cite{Silva:2016onh,Maas:2011ez} and
suggests that, similarly to the gluon propagator,
the ghost propagator can be used as an ``order parameter'' to identify the transition to the deconfinement region.
Note, however, that the increase in $d_G$ of about $\sim 1.2$ is smaller than that observed increase 
for the inverse of the electric (longitudinal) gluon propagator at zero momentum, almost a factor of $\sim 3$, when crossing $T_c$. 
Furthermore, note that our results are similar to previous results using quenched ensembles   
with smaller lattice volumes \cite{Aouane:2011fv}. However, our results suggest a stronger transition at $T_c$.

%====================================================================
%====================================================================
\section{Ghost Propagator and center symmetry \label{Sec:Ghost2}}

The generating functional of pure Yang-Mills theory on the lattice is invariant under center symmetry.  In the phase where
gluons are deconfined this symmetry is spontaneously broken. The breaking of the center symmetry can be identified by studying
the dynamics of the configurations associated with different phases of the Polyakov loop.

For the gluon propagator the effects associated with center symmetry, temperature and the breaking of the center symmetry 
were studied in \cite{Silva:2016onh}.  There, after separating the gauge configurations into different $Z_3$ sectors according to the 
phase of the Polyakov loop, the gluon propagator was computed in each of the sectors. 
It turns out that for temperatures below $T_c$ the gluon propagator is blind to the phase of the Polyakov loop. However,
for temperatures above $T_c$, i.e. for temperatures above the deconfinement temperature, the gluon propagator is no longer the same 
in all the sectors. Indeed, for $T > T_c$, the longitudinal (transverse) gluon propagator for the configurations that return a vanishing phase for the Polyakov loop 
is suppressed (enhanced) relative to the configurations where the phase of the Polyakov loop is $\pm \, 2 \, \pi /3$.  
This difference is an indication of the spontaneous breaking of the center symmetry that occurs at high temperatures. 
Moreover, the difference between the infrared gluon propagator associated with the $Z_3$ sectors can be used as an ``order parameter'' 
to describe the confinement-to-deconfinement transition. 
We now aim to investigate how the different dynamics, for temperatures below and above $T_c$, translate into the ghost propagator.

\begin{table}[t!]
\begin{center}
\begin{tabular}{ll@{\hspace{0.5cm}}l@{\hspace{0.5cm}}l@{\hspace{0.5cm}}l}
\hline
Temp.   & $L^3_s  \times L_t$ & $\beta$ & $a$  & $L_s a$ \\
 (MeV)  &                                 &              & (fm)  & (fm) \\
\hline
 270.1  & $72^3 \times 8$   &   6.058    & 0.09132 & 6.58  \\
 271.0  & $72^3 \times 8$   &   6.060    & 0.09101 & 6.55  \\
 271.5  & $72^3 \times 8$   &   6.061    & 0.09086 & 6.54 \\
 271.9  & $72^3 \times 8$   &   6.062    & 0.09071 & 6.53  \\
 272.4  & $72^3 \times 8$   &   6.063    & 0.09055 & 6.52  \\
 273.8  & $72^3 \times 8$   &   6.066    & 0.09010  & 6.49 \\
\hline
\end{tabular}
\end{center}
\caption{The lattice setup. The physical scale was defined from the string tension. 
              The values of $\beta$ were adjusted such that $L_s \, a \simeq 6.5 - 6.6$ fm.}
\label{tempsetup-center}
\end{table}

\begin{figure*}[t] %  figure placement: here, top, bottom, or page
   \centering
   \includegraphics[width=6in]{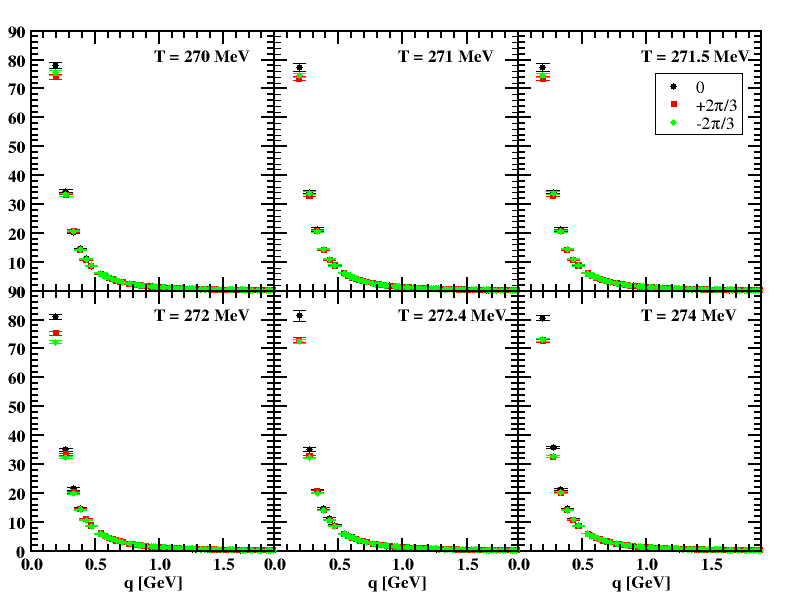}  \\
   \includegraphics[width=6in]{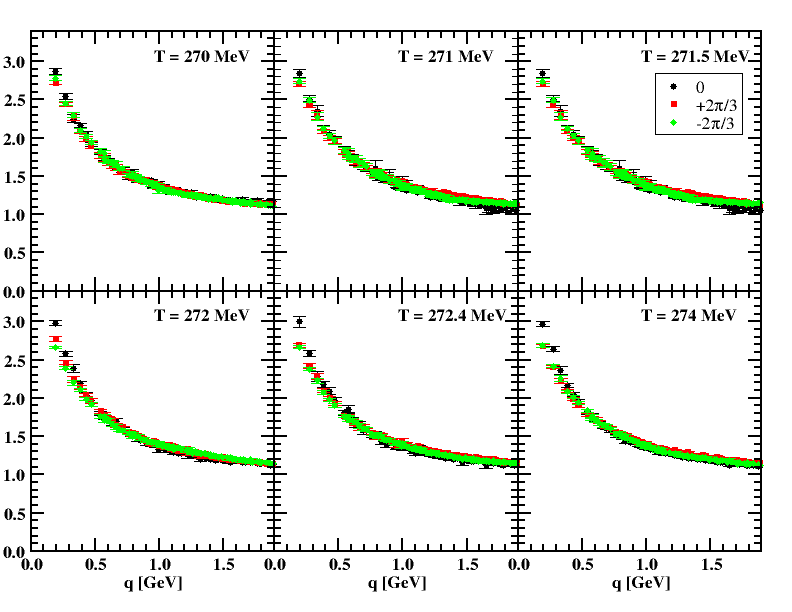} 
   \caption{Renormalized ghost propagator (top) and ghost dressing function (bottom) near $T_c$ and for each of the $Z_3$ sectors.}
   \label{fig:sectors}
\end{figure*}

In order to investigate center symmetry we consider the subset of configurations studied in \cite{Silva:2016onh} that are described in 
Tab. \ref{tempsetup-center}. 
In \cite{Silva:2016onh} and references therein, the reader can find the details about the characteristics of the gauge ensembles.
Note that the lattices are, in principle, sufficiently fine ($a \sim 0.1$fm)  and the  physical volumes are sufficiently large
($V \sim (6.5$ fm$)^4$) such that the finite volume effects are expected to be small \cite{Oliveira:2012eh}.

The ghost propagator and the ghost dressing function for temperatures near $T_c$ and for the various sectors defined according to the
phase of the Polyakov loop are reported in Fig. \ref{fig:sectors}. As can be seen, for the smallest temperatures the ghost propagator does
not depend on the phase of the Polyakov loop, while for the higher temperatures the ghost propagator for the zero sector is enhanced
relative to the propagators associated with the sectors whose Polyakov loop phase is $\pm \, 2 \, \pi / 3$. Note, however, that the propagators
associated with a $\pm \, 2 \, \pi / 3$ phase are indistinguishable. As  seen in \cite{Silva:2016onh}, the behavior observed for the ghost
propagator  follows that observed for the magnetic (transverse) form factor of the gluon propagator, as for the electric (longidutinal)
form factor the sector 0 is suppressed relative to the other two sectors that, once more, are indistinguishable. 

\begin{figure}[t] %  figure placement: here, top, bottom, or page
   \centering
   \includegraphics[width=3.4in]{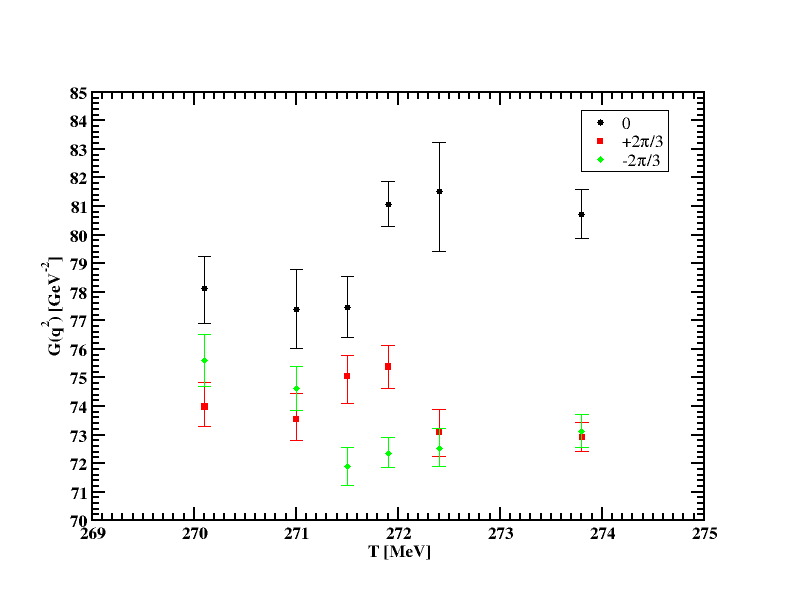}  
   % \caption{Renormalized ghost propagator  at the minimum computed $q = 0.19$ GeV near $T_c$ and for each of the $Z_3$ sectors.}
   \caption{Renormalized ghost propagator near $T_c$, at the smallest nonzero momentum $q = 0.19$ GeV  and for each of the $Z_3$ sectors.}
   \label{fig:sectors_min}
\end{figure}
 
 For completeness in Fig. \ref{fig:sectors_min} we report, for each of the $Z_3$ sectors, the ghost propagator at the  smallest non-zero momentum $q \sim 192$ MeV accessed in our  simulations.
 The Fig. shows that for the smaller temperatures the various propagators are compatible, within two standard deviations,
 while at the higher temperatures there is a clear separation between the data of the various propagators. Note that the $2\sigma$ separation for $T<T_c$ can be due to Gribov copies effects, something not explored in the current work.
 
 %============================================
 %============================================
 \section{Summary and Conclusions \label{Sec:fim}}
 
 In the current work the Landau gauge ghost propagator is studied at finite temperature with lattice simulations for pure Yang Mills SU(3) theory. 
Moreover, its relation with the breaking of the center symmetry is also addressed by looking at the dependence of the propagator with the phase of the Polyakov
loop.
 
 The simulations show an enhancement of the ghost form factor above the critical temperature $T_c$, 
 previously observed in  SU(3) studies on smaller volumes \cite{Aouane:2011fv}, while an  early investigation for the SU(2) gauge group 
 concluded in favour of a nearly independent ghost propagator with the temperature \cite{Cucchieri:2007ta}. 
 
 The behavior of the ghost propagator with respect to the center symmetry is also investigated and, similarly as found for the gluon propagator
 \cite{Silva:2016onh}, for $T < T_c$ the propagator for the various sectors are indistinguishable, while above $T_c$ the propagators associated
 with the $\pm 2 \pi/3$ sectors are suppressed when compared to the zero sector. Our results support the idea of using the difference of the ghost
 propagator as an ``order parameter'' to identify the transition to the deconfined phase. 
 
 We aim to extend this study over a wider range of temperatures and
 investigate the effect of the quark dynamics, measured through the quark propagator,  in each of the various $Z_3$ sectors
 using both the quenched  approximation and the full QCD \cite{Silva:2019cci}. We are also studyng the confinement-deconfinement
 transition in other gauges \cite{vanEgmond:2023lnw,empreparacao}.

 %============================================
 %============================================
 \section*{Acknowledgements}
 
 This work was partly supported by the FCT – Funda\c{c}\~ao para a Ci\^encia e a Tecnologia, I.P., under Projects Nos. UIDB/04564/2020, UIDP/04564/2020 and 
 CERN/FIS-COM/0029/2017. P. J. S. acknowledges financial support from FCT (Portugal) under Contract No. CEECIND/00488/2017. 
 The authors acknowledge the Laboratory for Advanced Computing at the University of Coimbra (http://www.uc.pt/lca) for providing access to the HPC resources
 that have contributed to the research within this paper.
 Access to Navigator was partly supported by the FCT Advanced Computing Project 2021.09759.CPCA. 

%%%%%%%%%%%%%%%%%%%%%%%%%   Bibliography   %%%%%%%%%%%%


\begin{thebibliography}{99}

%\cite{Fukushima:2010bq}
\bibitem{Fukushima:2010bq}
K.~Fukushima and T.~Hatsuda,
%``The phase diagram of dense QCD,''
Rept. Prog. Phys. \textbf{74}, 014001 (2011)
doi:10.1088/0034-4885/74/1/014001
[arXiv:1005.4814 [hep-ph]].
%824 citations counted in INSPIRE as of 15 Mar 2023

%\cite{Andersen:2014xxa}
\bibitem{Andersen:2014xxa}
J.~O.~Andersen, W.~R.~Naylor and A.~Tranberg,
%``Phase diagram of QCD in a magnetic field: A review,''
Rev. Mod. Phys. \textbf{88}, 025001 (2016)
doi:10.1103/RevModPhys.88.025001
[arXiv:1411.7176 [hep-ph]].
%311 citations counted in INSPIRE as of 15 Mar 2023

%\cite{Odyniec:2022ztn}
\bibitem{Odyniec:2022ztn}
G.~Odyniec,
%``Probing the QCD Phase Diagram with Heavy-Ion Collision Experiments,''
Lect. Notes Phys. \textbf{999}, 3-29 (2022)
doi:10.1007/978-3-030-95491-8\_1
%0 citations counted in INSPIRE as of 15 Mar 2023

%\cite{Oliveira:2018lln}
\bibitem{Oliveira:2018lln}
O.~Oliveira, P.~J.~Silva, J.~I.~Skullerud and A.~Sternbeck,
%``Quark propagator with two flavors of O(a)-improved Wilson fermions,''
Phys. Rev. D \textbf{99}, no.9, 094506 (2019)
doi:10.1103/PhysRevD.99.094506
[arXiv:1809.02541 [hep-lat]].
%46 citations counted in INSPIRE as of 14 Mar 2023

%\cite{Virgili:2022wfx}
\bibitem{Virgili:2022wfx}
A.~Virgili, W.~Kamleh and D.~Leinweber,
%``Overlap quark propagator near the physical pion mass,''
[arXiv:2209.14864 [hep-lat]].
%3 citations counted in INSPIRE as of 14 Mar 2023

%\cite{Oliveira:2019erx}
\bibitem{Oliveira:2019erx}
O.~Oliveira and P.~J.~Silva,
%``Finite Temperature Landau Gauge Lattice Quark Propagator,''
Eur. Phys. J. C \textbf{79}, no.9, 793 (2019)
doi:10.1140/epjc/s10052-019-7300-8
[arXiv:1903.00263 [hep-lat]].
%6 citations counted in INSPIRE as of 14 Mar 2023

%\cite{Borsanyi:2010bp}
\bibitem{Borsanyi:2010bp}
S.~Borsanyi \textit{et al.} [Wuppertal-Budapest],
%``Is there still any $T_c$ mystery in lattice QCD? Results with physical masses in the continuum limit III,''
JHEP \textbf{09}, 073 (2010)
doi:10.1007/JHEP09(2010)073
[arXiv:1005.3508 [hep-lat]].
%1083 citations counted in INSPIRE as of 15 Mar 2023

%\cite{Bazavov:2011nk}
\bibitem{Bazavov:2011nk}
A.~Bazavov, T.~Bhattacharya, M.~Cheng, C.~DeTar, H.~T.~Ding, S.~Gottlieb, R.~Gupta, P.~Hegde, U.~M.~Heller and F.~Karsch, \textit{et al.}
%``The chiral and deconfinement aspects of the QCD transition,''
Phys. Rev. D \textbf{85}, 054503 (2012)
doi:10.1103/PhysRevD.85.054503
[arXiv:1111.1710 [hep-lat]].
%1196 citations counted in INSPIRE as of 15 Mar 2023

%\cite{Dumitru:2012fw}
\bibitem{Dumitru:2012fw}
A.~Dumitru, Y.~Guo, Y.~Hidaka, C.~P.~K.~Altes and R.~D.~Pisarski,
%``Effective Matrix Model for Deconfinement in Pure Gauge Theories,''
Phys. Rev. D \textbf{86}, 105017 (2012)
doi:10.1103/PhysRevD.86.105017
[arXiv:1205.0137 [hep-ph]].
%95 citations counted in INSPIRE as of 14 Mar 2023

%\cite{Maas:2011ez}
\bibitem{Maas:2011ez}
A.~Maas, J.~M.~Pawlowski, L.~von Smekal and D.~Spielmann,
%``The Gluon propagator close to criticality,''
Phys. Rev. D \textbf{85}, 034037 (2012)
doi:10.1103/PhysRevD.85.034037
[arXiv:1110.6340 [hep-lat]].
%83 citations counted in INSPIRE as of 14 Mar 2023

%\cite{Aouane:2011fv}
\bibitem{Aouane:2011fv}
R.~Aouane, V.~G.~Bornyakov, E.~M.~Ilgenfritz, V.~K.~Mitrjushkin, M.~Muller-Preussker and A.~Sternbeck,
%``Landau gauge gluon and ghost propagators at finite temperature from quenched lattice QCD,''
Phys. Rev. D \textbf{85}, 034501 (2012)
doi:10.1103/PhysRevD.85.034501
[arXiv:1108.1735 [hep-lat]].
%83 citations counted in INSPIRE as of 14 Mar 2023

%\cite{Bornyakov:2011jm}
\bibitem{Bornyakov:2011jm}
V.~G.~Bornyakov and V.~K.~Mitrjushkin,
%``Lattice QCD gluon propagators near transition temperature,''
Int. J. Mod. Phys. A \textbf{27}, 1250050 (2012)
doi:10.1142/S0217751X12500509
[arXiv:1103.0442 [hep-lat]].
%32 citations counted in INSPIRE as of 14 Mar 2023

%\cite{Maas:2011se}
\bibitem{Maas:2011se}
A.~Maas,
%``Describing gauge bosons at zero and finite temperature,''
Phys. Rept. \textbf{524}, 203-300 (2013)
doi:10.1016/j.physrep.2012.11.002
[arXiv:1106.3942 [hep-ph]].
%255 citations counted in INSPIRE as of 14 Mar 2023

%\cite{Aouane:2012bk}
\bibitem{Aouane:2012bk}
R.~Aouane, F.~Burger, E.~M.~Ilgenfritz, M.~M\"uller-Preussker and A.~Sternbeck,
%``Landau gauge gluon and ghost propagators from lattice QCD with $N_f$=2 twisted mass fermions at finite temperature,''
Phys. Rev. D \textbf{87}, no.11, 114502 (2013)
doi:10.1103/PhysRevD.87.114502
[arXiv:1212.1102 [hep-lat]].
%42 citations counted in INSPIRE as of 15 Mar 2023

%\cite{Silva:2013maa}
\bibitem{Silva:2013maa}
P.~J.~Silva, O.~Oliveira, P.~Bicudo and N.~Cardoso,
%``Gluon screening mass at finite temperature from the Landau gauge gluon propagator in lattice QCD,''
Phys. Rev. D \textbf{89}, no.7, 074503 (2014)
doi:10.1103/PhysRevD.89.074503
[arXiv:1310.5629 [hep-lat]].
%72 citations counted in INSPIRE as of 14 Mar 2023

%\cite{Falcao:2020vyr}
\bibitem{Falcao:2020vyr}
A.~F.~Falc\~ao, O.~Oliveira and P.~J.~Silva,
%``Analytic structure of the lattice Landau gauge gluon and ghost propagators,''
Phys. Rev. D \textbf{102}, no.11, 114518 (2020)
doi:10.1103/PhysRevD.102.114518
[arXiv:2008.02614 [hep-lat]].
%33 citations counted in INSPIRE as of 15 Mar 2023

%\cite{Siringo:2021fxo}
\bibitem{Siringo:2021fxo}
F.~Siringo and G.~Comitini,
%``Thermal extension of the screened massive expansion in the Landau gauge,''
Phys. Rev. D \textbf{103}, no.7, 074014 (2021)
doi:10.1103/PhysRevD.103.074014
[arXiv:2101.08341 [hep-th]].
%5 citations counted in INSPIRE as of 15 Mar 2023

%\cite{vanEgmond:2022nuo}
\bibitem{vanEgmond:2022nuo}
D.~M.~van Egmond and U.~Reinosa,
%``Signatures of the Yang-Mills deconfinement transition from the gluon two-point correlator,''
Phys. Rev. D \textbf{106}, no.7, 074005 (2022)
doi:10.1103/PhysRevD.106.074005
[arXiv:2206.03841 [hep-ph]].
%1 citations counted in INSPIRE as of 14 Mar 2023

%\cite{Cornwall:2013zra}
\bibitem{Cornwall:2013zra}
J.~M.~Cornwall,
%``Positivity violations in QCD,''
Mod. Phys. Lett. A \textbf{28}, 1330035 (2013)
doi:10.1142/S0217732313300358
[arXiv:1310.7897 [hep-ph]].
%48 citations counted in INSPIRE as of 15 Mar 2023

%\cite{Li:2019hyv}
\bibitem{Li:2019hyv}
S.~W.~Li, P.~Lowdon, O.~Oliveira and P.~J.~Silva,
%``The generalised infrared structure of the gluon propagator,''
Phys. Lett. B \textbf{803}, 135329 (2020)
doi:10.1016/j.physletb.2020.135329
[arXiv:1907.10073 [hep-th]].
%27 citations counted in INSPIRE as of 15 Mar 2023

%\cite{Oliveira:2022rwu}
\bibitem{Oliveira:2022rwu}
O.~Oliveira, L.~C.~Loveridge and P.~J.~Silva,
%``Compact QED: the photon propagator, confinement and positivity violation for the pure gauge theory,''
EPJ Web Conf. \textbf{274}, 02004 (2022)
doi:10.1051/epjconf/202227402004
[arXiv:2211.12593 [hep-lat]].
%0 citations counted in INSPIRE as of 15 Mar 2023

%\cite{Kapusta:2006pm}
\bibitem{Kapusta:2006pm}
J.~I.~Kapusta and C.~Gale,
%``Finite-temperature field theory: Principles and applications,''
Cambridge University Press, 2011,
ISBN 978-0-521-17322-3, 978-0-521-82082-0, 978-0-511-22280-1
doi:10.1017/CBO9780511535130
%273 citations counted in INSPIRE as of 15 Mar 2023

%\cite{Silva:2016onh}
\bibitem{Silva:2016onh}
P.~J.~Silva and O.~Oliveira,
%``Gluon Dynamics, Center Symmetry and the deconfinement phase transition in SU(3) pure Yang-Mills theory,''
Phys. Rev. D \textbf{93}, no.11, 114509 (2016)
doi:10.1103/PhysRevD.93.114509
[arXiv:1601.01594 [hep-lat]].
%21 citations counted in INSPIRE as of 15 Mar 2023

%\cite{Lo:2013etb}
\bibitem{Lo:2013etb}
P.~M.~Lo, B.~Friman, O.~Kaczmarek, K.~Redlich and C.~Sasaki,
%``Probing Deconfinement with Polyakov Loop Susceptibilities,''
Phys. Rev. D \textbf{88}, no.1, 014506 (2013)
doi:10.1103/PhysRevD.88.014506
[arXiv:1306.5094 [hep-lat]].
%43 citations counted in INSPIRE as of 15 Mar 2023

%\cite{Cucchieri:1997dx}
\bibitem{Cucchieri:1997dx}
A.~Cucchieri,
%``Gribov copies in the minimal Landau gauge: The Influence on gluon and ghost propagators,''
Nucl. Phys. B \textbf{508}, 353-370 (1997)
doi:10.1016/S0550-3213(97)00629-9
[arXiv:hep-lat/9705005 [hep-lat]].
%164 citations counted in INSPIRE as of 15 Mar 2023

%\cite{Silva:2007tt}
\bibitem{Silva:2007tt}
P.~J.~Silva and O.~Oliveira,
%``Gauge fixing methods and Gribov copies effects in lattice QCD,''
PoS \textbf{LATTICE2007}, 333 (2007)
doi:10.22323/1.042.0333
[arXiv:0710.0669 [hep-lat]].
%6 citations counted in INSPIRE as of 15 Mar 2023

%\cite{Silva:2010vx}
\bibitem{Silva:2010vx}
P.~J.~Silva and O.~Oliveira,
%``Unquenching the Landau Gauge Lattice Propagators and the Gribov Problem,''
PoS \textbf{LATTICE2010}, 287 (2010)
doi:10.22323/1.105.0287
[arXiv:1011.0483 [hep-lat]].
%6 citations counted in INSPIRE as of 15 Mar 2023

%\cite{Silva:2004bv}
\bibitem{Silva:2004bv}
P.~J.~Silva and O.~Oliveira,
%``Gribov copies, lattice QCD and the gluon propagator,''
Nucl. Phys. B \textbf{690}, 177-198 (2004)
doi:10.1016/j.nuclphysb.2004.04.020
[arXiv:hep-lat/0403026 [hep-lat]].
%104 citations counted in INSPIRE as of 15 Mar 2023

%\cite{Sternbeck:2012mf}
\bibitem{Sternbeck:2012mf}
A.~Sternbeck and M.~M\"uller-Preussker,
%``Lattice evidence for the family of decoupling solutions of Landau gauge Yang-Mills theory,''
Phys. Lett. B \textbf{726}, 396-403 (2013)
doi:10.1016/j.physletb.2013.08.017
[arXiv:1211.3057 [hep-lat]].
%74 citations counted in INSPIRE as of 15 Mar 2023


%\cite{Kaiser:2022vjg}
\bibitem{Kaiser:2022vjg}
R.~Kaiser and O.~Philipsen,
%``Progress on the QCD Deconfinement Critical Point for $N_\text{f}=2$ Staggered Fermions,''
PoS \textbf{LATTICE2022}, 175 (2023)
doi:10.22323/1.430.0175
[arXiv:2212.14461 [hep-lat]].
%1 citations counted in INSPIRE as of 17 Mar 2023

%\cite{Oliveira:2022jir}
\bibitem{Oliveira:2022jir}
O.~Oliveira, V.~Paiva and P.~Silva,
%``Deconfinement in pure gauge SU(3) Yang-Mills theory: the ghost propagator,''
EPJ Web Conf. \textbf{274}, 05008 (2022)
doi:10.1051/epjconf/202227405008
[arXiv:2301.01229 [hep-lat]].
%0 citations counted in INSPIRE as of 17 Mar 2023

%\cite{Suman:1995zg}
\bibitem{Suman:1995zg}
H.~Suman and K.~Schilling,
%``First lattice study of ghost propagators in SU(2) and SU(3) gauge theories,''
Phys. Lett. B \textbf{373}, 314-318 (1996)
doi:10.1016/0370-2693(96)00162-1
[arXiv:hep-lat/9512003 [hep-lat]].
%86 citations counted in INSPIRE as of 04 Apr 2023

%\cite{Leinweber:1998uu}
\bibitem{Leinweber:1998uu}
D.~B.~Leinweber \textit{et al.} [UKQCD],
%``Asymptotic scaling and infrared behavior of the gluon propagator,''
Phys. Rev. D \textbf{60}, 094507 (1999)
[erratum: Phys. Rev. D \textbf{61}, 079901 (2000)]
doi:10.1103/PhysRevD.60.094507
[arXiv:hep-lat/9811027 [hep-lat]].
%209 citations counted in INSPIRE as of 04 Apr 2023

%\cite{Montvay:1994cy}
\bibitem{Montvay:1994cy}
I.~Montvay and G.~Munster,
%``Quantum fields on a lattice,''
Cambridge University Press, 1997,
ISBN 978-0-521-59917-7, 978-0-511-87919-7
doi:10.1017/CBO9780511470783
%152 citations counted in INSPIRE as of 05 Apr 2023

%\cite{Becirevic:1999uc}
\bibitem{Becirevic:1999uc}
D.~Becirevic, P.~Boucaud, J.~P.~Leroy, J.~Micheli, O.~Pene, J.~Rodriguez-Quintero and C.~Roiesnel,
%``Asymptotic behavior of the gluon propagator from lattice QCD,''
Phys. Rev. D \textbf{60}, 094509 (1999)
doi:10.1103/PhysRevD.60.094509
[arXiv:hep-ph/9903364 [hep-ph]].
%108 citations counted in INSPIRE as of 05 Apr 2023



%\cite{deSoto:2007ht}
\bibitem{deSoto:2007ht}
F.~de Soto and C.~Roiesnel,
%``On the reduction of hypercubic lattice artifacts,''
JHEP \textbf{09}, 007 (2007)
doi:10.1088/1126-6708/2007/09/007
[arXiv:0705.3523 [hep-lat]].
%77 citations counted in INSPIRE as of 05 Apr 2023

%\cite{Vujinovic:2018nqc}
\bibitem{Vujinovic:2018nqc}
M.~Vujinovic and T.~Mendes,
%``Probing the tensor structure of lattice three-gluon vertex in Landau gauge,''
Phys. Rev. D \textbf{99}, no.3, 034501 (2019)
doi:10.1103/PhysRevD.99.034501
[arXiv:1807.03673 [hep-lat]].
%21 citations counted in INSPIRE as of 05 Apr 2023

%\cite{Catumba:2021hcx}
\bibitem{Catumba:2021hcx}
G.~T.~R.~Catumba, O.~Oliveira and P.~J.~Silva,
%``$H(4)$ tensor representations for the lattice Landau gauge gluon propagator and the estimation of lattice artefacts,''
Phys. Rev. D \textbf{103}, no.7, 074501 (2021)
doi:10.1103/PhysRevD.103.074501
[arXiv:2101.04978 [hep-lat]].
%8 citations counted in INSPIRE as of 05 Apr 2023

%\cite{Cucchieri:2007md}
\bibitem{Cucchieri:2007md}
A.~Cucchieri and T.~Mendes,
%``What's up with IR gluon and ghost propagators in Landau gauge? A puzzling answer from huge lattices,''
PoS \textbf{LATTICE2007}, 297 (2007)
doi:10.22323/1.042.0297
[arXiv:0710.0412 [hep-lat]].
%386 citations counted in INSPIRE as of 05 Apr 2023

%\cite{Cucchieri:2007zm}
\bibitem{Cucchieri:2007zm}
A.~Cucchieri, T.~Mendes, O.~Oliveira and P.~J.~Silva,
%``Just how different are SU(2) and SU(3) Landau propagators in the IR regime?,''
Phys. Rev. D \textbf{76}, 114507 (2007)
doi:10.1103/PhysRevD.76.114507
[arXiv:0705.3367 [hep-lat]].
%75 citations counted in INSPIRE as of 05 Apr 2023

%\cite{Bogolubsky:2009dc}
\bibitem{Bogolubsky:2009dc}
I.~L.~Bogolubsky, E.~M.~Ilgenfritz, M.~Muller-Preussker and A.~Sternbeck,
%``Lattice gluodynamics computation of Landau gauge Green's functions in the deep infrared,''
Phys. Lett. B \textbf{676}, 69-73 (2009)
doi:10.1016/j.physletb.2009.04.076
[arXiv:0901.0736 [hep-lat]].
%499 citations counted in INSPIRE as of 05 Apr 2023

%\cite{Ilgenfritz:2010gu}
\bibitem{Ilgenfritz:2010gu}
E.~M.~Ilgenfritz, C.~Menz, M.~Muller-Preussker, A.~Schiller and A.~Sternbeck,
%``SU(3) Landau gauge gluon and ghost propagators using the logarithmic lattice gluon field definition,''
Phys. Rev. D \textbf{83}, 054506 (2011)
doi:10.1103/PhysRevD.83.054506
[arXiv:1010.5120 [hep-lat]].
%28 citations counted in INSPIRE as of 05 Apr 2023

%\cite{Duarte:2016iko}
\bibitem{Duarte:2016iko}
A.~G.~Duarte, O.~Oliveira and P.~J.~Silva,
%``Lattice Gluon and Ghost Propagators, and the Strong Coupling in Pure SU(3) Yang-Mills Theory: Finite Lattice Spacing and Volume Effects,''
Phys. Rev. D \textbf{94}, no.1, 014502 (2016)
doi:10.1103/PhysRevD.94.014502
[arXiv:1605.00594 [hep-lat]].
%109 citations counted in INSPIRE as of 05 Apr 2023

%\cite{Dudal:2019gvn}
\bibitem{Dudal:2019gvn}
D.~Dudal, O.~Oliveira, M.~Roelfs and P.~Silva,
%``Spectral representation of lattice gluon and ghost propagators at zero temperature,''
Nucl. Phys. B \textbf{952}, 114912 (2020)
doi:10.1016/j.nuclphysb.2019.114912
[arXiv:1901.05348 [hep-lat]].
%43 citations counted in INSPIRE as of 05 Apr 2023

%\cite{Boito:2022rad}
\bibitem{Boito:2022rad}
D.~Boito, A.~Cucchieri, C.~Y.~London and T.~Mendes,
%``Probing the singularities of the Landau-Gauge gluon and ghost propagators with rational approximants,''
JHEP \textbf{02}, 144 (2023)
doi:10.1007/JHEP02(2023)144
[arXiv:2210.10490 [hep-lat]].
%5 citations counted in INSPIRE as of 05 Apr 2023

%\cite{Duarte:2016ieu}
\bibitem{Duarte:2016ieu}
A.~G.~Duarte, O.~Oliveira and P.~J.~Silva,
%``Further Evidence For Zero Crossing On The Three Gluon Vertex,''
Phys. Rev. D \textbf{94}, no.7, 074502 (2016)
doi:10.1103/PhysRevD.94.074502
[arXiv:1607.03831 [hep-lat]].
%53 citations counted in INSPIRE as of 05 Apr 2023

%\cite{Iwasaki:1992ik}
\bibitem{Iwasaki:1992ik}
Y.~Iwasaki, K.~Kanaya, T.~Yoshie, T.~Hoshino, T.~Shirakawa, Y.~Oyanagi, S.~Ichii and T.~Kawai,
%``Finite temperature phase transition of SU(3) gauge theory on N(t) = 4 and 6 lattices,''
Phys. Rev. D \textbf{46}, 4657-4667 (1992)
doi:10.1103/PhysRevD.46.4657
%87 citations counted in INSPIRE as of 05 Apr 2023

%\cite{Boyd:1996bx}
\bibitem{Boyd:1996bx}
G.~Boyd, J.~Engels, F.~Karsch, E.~Laermann, C.~Legeland, M.~Lutgemeier and B.~Petersson,
%``Thermodynamics of SU(3) lattice gauge theory,''
Nucl. Phys. B \textbf{469}, 419-444 (1996)
doi:10.1016/0550-3213(96)00170-8
[arXiv:hep-lat/9602007 [hep-lat]].
%1122 citations counted in INSPIRE as of 05 Apr 2023

%\cite{Lucini:2001ej}
\bibitem{Lucini:2001ej}
B.~Lucini and M.~Teper,
%``SU(N) gauge theories in four-dimensions: Exploring the approach to N = infinity,''
JHEP \textbf{06}, 050 (2001)
doi:10.1088/1126-6708/2001/06/050
[arXiv:hep-lat/0103027 [hep-lat]].
%244 citations counted in INSPIRE as of 05 Apr 2023

%\cite{Maas:2011ez}
\bibitem{Maas:2011ez}
A.~Maas, J.~M.~Pawlowski, L.~von Smekal and D.~Spielmann,
%``The Gluon propagator close to criticality,''
Phys. Rev. D \textbf{85}, 034037 (2012)
doi:10.1103/PhysRevD.85.034037
[arXiv:1110.6340 [hep-lat]].
%84 citations counted in INSPIRE as of 05 Apr 2023

%\cite{Oliveira:2012eh}
\bibitem{Oliveira:2012eh}
O.~Oliveira and P.~J.~Silva,
%``The lattice Landau gauge gluon propagator: lattice spacing and volume dependence,''
Phys. Rev. D \textbf{86}, 114513 (2012)
doi:10.1103/PhysRevD.86.114513
[arXiv:1207.3029 [hep-lat]].
%149 citations counted in INSPIRE as of 10 Apr 2023


\bibitem{chroma}
R. G. Edwards et al. [SciDAC, LHPC and UKQCD],
Nucl. Phys. B Proc. Suppl. 140, 832 (2005)
doi:10.1016/j.nuclphysbps.2004.11.254 [arXiv:heplat/0409003 [hep-lat]].

%\cite{Cucchieri:2007ta}
\bibitem{Cucchieri:2007ta}
A.~Cucchieri, A.~Maas and T.~Mendes,
%``Infrared properties of propagators in Landau-gauge pure Yang-Mills theory at finite temperature,''
Phys. Rev. D \textbf{75}, 076003 (2007)
doi:10.1103/PhysRevD.75.076003
[arXiv:hep-lat/0702022 [hep-lat]].
%115 citations counted in INSPIRE as of 31 May 2023

%\cite{Silva:2019cci}
\bibitem{Silva:2019cci}
P.~J.~Silva and O.~Oliveira,
%``Lattice computation of the quark propagator in Landau gauge at finite temperature,''
PoS \textbf{LATTICE2019}, 047 (2020)
doi:10.22323/1.363.0047
[arXiv:1912.13061 [hep-lat]].
%1 citations counted in INSPIRE as of 17 Jul 2023

\bibitem{pfft}
M. Pippig, SIAM J. Sci. Comput. 35, C213 (2013)

%\cite{vanEgmond:2023lnw}
\bibitem{vanEgmond:2023lnw}
D.~M.~van Egmond and U.~Reinosa,
%``Gauge fixing and physical symmetries,''
[arXiv:2304.00756 [hep-th]].
%0 citations counted in INSPIRE as of 21 Jun 2023

%\cite{vanEgmond:2023lnw}
\bibitem{empreparacao}
D.~M.~van Egmond, U.~Reinosa, O. Oliveira, P. J. Silva,
\textit{in preparation}

\end{thebibliography}
\end{document}